\documentclass[10pt,conference]{IEEEtran}

\usepackage[dvipsnames]{xcolor}
\usepackage{amsthm}
\usepackage{thmtools}
\usepackage{mathtools}
\usepackage{bm} % bold math
\usepackage{bbm} % black board math
\usepackage{braket}
\usepackage{dsfont}
\usepackage{amsmath}
\usepackage{amssymb}
\usepackage{tikz}
\usepackage[most]{tcolorbox}
\usetikzlibrary{arrows, positioning}
\usepackage{tikz-cd}
\usepackage{graphicx}

\usepackage[linesnumbered,ruled]{algorithm2e}
\SetArgSty{textnormal}

\usepackage{etoolbox}
\apptocmd{\sloppy}{\hbadness 10000\relax}{}{}

\usepackage{silence}
    \DeclarePairedDelimiter{\abs}{\lvert}{\rvert} % absolute value
        \DeclareMathOperator{\KP}{KP} %knapsack problem
        \DeclareMathOperator*{\bin}{bin} % binary representation
        \DeclareMathOperator*{\spn}{span} % linear span
        \newcommand*{\N}{\mathbb{N}}

        \newcommand*{\R}{\mathbb{R}}

        \newcommand*{\idty}{\mathds{1}}
        \newcommand*{\defcolon}{\,:\,} % colon for e.g. definition in sets
        \newcommand*{\numbits}[1]{\abs{{#1}_{\bin}}} % length of binary representation
        \newcommand*{\scriptin}{\raisebox{0.15ex}{$\scriptscriptstyle\in$}} % "element of" symbol suited for sub- and superscripts
        \newcommand*{\expval}[3]{\bra{#1}\negthickspace#2\negthickspace\ket{#3}} % expectation value
        \newcommand*{\ketbra}[2]{\ket{#1}\negthickspace\bra{#2}} % ket-bra notation
        \newcommand*{\ketbraket}[3]{\ket{#1}\negthickspace\braket{#2 | #3}} % ket-bra-ket notation
        
            \newcommand*{\kp}{\KP_{n}(\bm{v}, \bm{w}; c)} % generic knapsack instance
            \newcommand*{\feasset}{S} % feasible set
            \newcommand*{\feasspace}{\langle \feasset\rangle} % feasible subspace
    
            \newcommand*{\phasesep}{U_{\text{P}}} % phase separator
            \newcommand*{\mixer}{U_{\text{M}}} % mixer
            \newcommand*{\stateprep}{U_{\text{SP}}} % state preparation
        
            \newcommand*{\QTG}{\mathcal{G}} % state preparation circuit
            \newcommand*{\controlled}{C} % control
            \newcommand*{\qubitc}{\ensuremath{\mathbf{Q}}} % qubit count
            \newcommand*{\cyclec}{\ensuremath{\mathbf{C}}} % cycle count

\usepackage[colorlinks]{hyperref}
\hypersetup{
    colorlinks  = true,
    citecolor   = PineGreen,
    linkcolor   = MidnightBlue,
    urlcolor    = PineGreen
}

\makeatletter
\def\ps@IEEEtitlepagestyle{%
  \def\@oddfoot{\mycopyrightnotice}%
  \def\@evenfoot{}%
}
\def\mycopyrightnotice{%
  {\footnotesize
  \hfill 
  \parbox{\textwidth}{%
  © 2025 IEEE.  Personal use of this material is permitted.  Permission from IEEE must be obtained for all other uses, in any current or future media, including reprinting/republishing this material for advertising or promotional purposes, creating new collective works, for resale or redistribution to servers or lists, or reuse of any copyrighted component of this work in other works.}
  \hfill}
}
\makeatother

\begin{document}

\title{Quantum tree generator improves QAOA state-of-the-art for the knapsack problem}

\author{\IEEEauthorblockN{Paul J. Christiansen\IEEEauthorrefmark{1}${}^{,}$\IEEEauthorrefmark{2}, Lennart Binkowski\IEEEauthorrefmark{2}, Debora Ramacciotti\IEEEauthorrefmark{2}, S\"oren Wilkening\IEEEauthorrefmark{2}}
\IEEEauthorblockA{\IEEEauthorrefmark{1}d-fine GmbH, Frankfurt, Germany\\
Email: paul.christiansen@d-fine.com}
\IEEEauthorblockA{\IEEEauthorrefmark{2}Institut f\"ur Theoretische Physik\\
Leibniz Universit\"at Hannover, Hannover, Germany}}

\maketitle

\begin{abstract}
    This paper introduces a novel approach to the Quantum Approximate Optimization Algorithm (QAOA), specifically tailored to the knapsack problem.
    We combine the recently proposed quantum tree generator as an efficient state preparation circuit for all feasible solutions to the knapsack problem with the framework of Grover-mixer QAOA to form the first representative of Amplitude Amplification-mixer QAOA (AAM-QAOA).
    On hard benchmark sets with up to 20 knapsack items, we demonstrate our method's improved performance over the current state-of-the-art Copula-QAOA.
    However, for larger instance sizes, both approaches fail to deliver better outcomes than greedily packing items in descending value-to-weight ratio, at least for the considered circuit depths.
    For sufficiently high circuit depths, however, we can prove that AAM-QAOA will eventually be able to sample the optimal solution.
\end{abstract}


\section{Introduction}\label{section:Introduction}

Within the current era of noisy intermediate-scale quantum (NISQ) computers~\cite{Preskill2018QuantumComputingInTheNISQEraAndBeyond}, variational quantum algorithms~\cite{Cerezo2021VariationalQuantumAlgorithms} constitute the most promising family of hybrid algorithms for supplying quantum advantage over purely classical methods.
Especially in the domain of quantum simulation~\cite{Georgescu2014QuantumSimulation} and related tasks, classical approaches seem to face fundamental difficulties that are believed to be (at least partially) overcome by utilizing the promising capabilities of quantum computers.
Similarly, quantum computers and algorithms have also been considered as potential accelerators for NP-hard classical problems, such as most combinatorial optimization problem formulations (maximize $f : \{0, 1\}^{n} \rightarrow \R$ subject to some constraints).
The \emph{quantum approximate optimization algorithm}~\cite{Farhi2014AQuantumApproximateOptimizationAlgorithm} and its generalization to the \emph{quantum alternating operator ansatz}~\cite{Hadfield2019FromTheQuantumApproximateOptimizationAlgorithmToAQuantumAlternatingOperatorAnsatz} (QAOA) are arguably the most influential variational quantum algorithms for these kinds of tasks.
In a nutshell, QAOA aims at maximizing the expectation value of the Hamiltonian encoding the classical objective function $f$ over a parameterized class of quantum states.
This class is realized by applying parameterized quantum circuits to a given fixed initial state.
The parameter optimization is outsourced to a classical optimizer, while the quantum computer is mainly used as a sampling machine, providing fast evaluations/estimations of the objective function.
For constrained problems, there are several approaches to address the problems' feasibility structure.
One possibility is to augment the objective function with soft-constraints, penalizing infeasible states such that feasible solutions are favored, i.e.\ yield a higher expectation value.
Alternatively, one may actively restrict the parameterized class of states to the feasible subspace (the space spanned by all feasible solutions) by selecting a feasible initial state and feasibility-preserving quantum circuits, that is, circuits that leave the feasible subspace invariant.
In this case, soft-constraints are redundant, as only feasible states can be sampled in the first place.
However, this approach usually requires additional insights regarding the problem's feasibility structure -- knowledge that might not always be available.

The 0-1 knapsack problem (0-1-KP) is a prototypical combinatorial optimization problem, whose feasibility structure is difficult to access:
Given a set of items, each having a profit and a weight, and a knapsack of certain capacity, the task is to maximize the cumulative profit of items placed in the knapsack while keeping their cumulative weight below or equal to the capacity.
Despite the compact description of its constraint, there is no apparent symmetry in the 0-1-KP's feasible set.
This makes it particularly difficult to construct feasibility-preserving quantum circuits.
Accordingly, most QAOA formulations for the 0-1-KP are based on soft-constraints \cite{Lucas2014IsingFormulationOfManyNPProblems,DeLaGrandrive2019KnapsackProblemVariantsOfQAOAForBatteryRevenueOptimisation,Roch2021CrossEntropyOptimizationOfConstrainedProblemHamiltoniansForQuantumAnnealing}.
However, this approach comes with some caveats such:
Choosing the penalty too high may blanket the optimization landscape;
choosing it too low may lead to significant feasibility violation.
With the recently proposed Copula-QAOA, van Dam \textit{et al.}~\cite{VanDam2021QuantumOptimizationHeuristicsWithAnApplicationToKnapsackProblems} resolve this issue of fine-tuning the penalty by simply assigning an objective value of zero to all infeasible states.
Furthermore, they implement the idea of warm-starting QAOA~\cite{Egger2021WarmStartingQuantumOptimization} with an efficient classical algorithm specifically for the 0-1-KP.
Their numerical experiments on instances up to ten qubits look promising;
in this work we extend benchmarking their approach for higher circuit depths and harder instances with up to 20 qubits as provided by \cite{Jooken2022ANewClassOfHardProblemInstancesForTheKnapsackProblem} \textit{et al.}'s instance generator.
Unfortunately, we observe a significant decrease in solution quality, for instance with more than ten qubits.

As an alternative, we propose to embed the recently proposed \emph{Quantum Tree Generator} (QTG)~\cite{Wilkening2023AQuantumAlgorithmForTheSolutionOfTheKnapsackProblem} into the framework of \emph{Grover-mixer QAOA} (GM-QAOA)~\cite{Baertschi2020GroverMixersForQAOAShiftingComplexityFromMixerDesignToStatePreparation}.
Given a state preparation circuit that creates a uniform superposition of all feasible states, GM-QAOA prescribes using a parameterized version of that circuit in order to construct a fully feasible parameterized class of states.
The QTG, in turn, is an efficient quantum circuit that creates a generally non-uniform superposition of all feasible solutions to the 0-1-KP.
Utilizing it instead of a uniform state preparation should not have any drawback.
Indeed, we are able to prove that QAOA with the QTG as Grover-like mixer converges towards the optimum as the circuit depth increases.
Our proof is a direct application of the convergence theorem provided in~\cite{Binkowski2024ElementaryProofOfQAOAConvergence}.
Furthermore, we conduct numerical experiments for our method on the same benchmark instances as for the Copula-QAOA and demonstrate that our method is able to provide higher approximation ratios on a consistent level, that is, without the drop in quality for instance sizes above ten.
By adapting the high-level simulator presented in~\cite{Wilkening2023AQuantumAlgorithmForTheSolutionOfTheKnapsackProblem}, we are able to extend our simulations for the QTG-based QAOA for instances with up to 34 items.

\section{Previous work}\label{section:PreviousWork}

On the classical side, there exist several prominent algorithms, specifically formulated to solve instances of the 0-1-KP.
Already in 1995, Pisinger~\cite{Pisinger1995AnExpandingCoreAlgorithmForTheExactKnapsackProblem} developed the EXPKNAP algorithm, a 0-1-KP solver based on branch-and-bound methods.
In 1999, EXPKNAP and subsequent refinements were superseded by the dynamic-programming-based COMBO algorithm~\cite{Martello1999DynamicProgrammingAndStrongBoundsForTheKnapsackProblem} which still remains the best performing candidate across various benchmark sets.
Beside these specialized algorithms, also more general frameworks based on integer programming and constraint programming such as CPLEX~\cite{IBM2022IBMILOGCPLEXOptimizationStudio}, GUROBI~\cite{GurobiOptimizationLLC2023GurobiOptimizerReferenceManual}, or CP-SAT~\cite{GoogleORTools2024GoogleORToolsDocumentation} may be applied to instances of 0-1-KP.
For our numerical studies, we merely utilize the COMBO method in order to extract crucial data of the benchmark instances;
our own contributions are conceptually disjoint with all the just mentioned algorithms.

On the quantum-algorithmic side, 0-1-KP is frequently featured within the scope of heuristic approaches such as QAOA (see e.g.\ \cite{Lucas2014IsingFormulationOfManyNPProblems,DeLaGrandrive2019KnapsackProblemVariantsOfQAOAForBatteryRevenueOptimisation,Roch2021CrossEntropyOptimizationOfConstrainedProblemHamiltoniansForQuantumAnnealing}).
Most prominently, van Dam \textit{et al.}~\cite{VanDam2021QuantumOptimizationHeuristicsWithAnApplicationToKnapsackProblems} recently introduced a version of the QAOA specifically tailored to 0-1-KP.
Further prominent enhancements of the QAOA without a direct link to 0-1-KP are the quantum alternating operator ansatz~\cite{Hadfield2019FromTheQuantumApproximateOptimizationAlgorithmToAQuantumAlternatingOperatorAnsatz} and Grover mixer-QAOA~\cite{Baertschi2020GroverMixersForQAOAShiftingComplexityFromMixerDesignToStatePreparation}.
Beside expanding on the capabilities of van Dam \textit{et al.}'s approach, we primarily establish that missing link between 0-1-KP and the just mentioned QAOA extensions.

In contrast to near-term heuristics such as the QAOA, several prominent quantum frameworks such as nested quantum search~\cite{Cerf2000NestedQuantumSearchAndStructuredProblems} and quantum branch-and-bound~\cite{Montanaro2020QuantumSpeedupOfBranchAndBoundAlgorithms} are suitable candidates for far-term quantum methods that could be applied to instances of 0-1-KP.
A far-term approach specifically tailored to 0-1-KP was recently proposed by Wilkening \textit{et al.}~
\cite{Wilkening2023AQuantumAlgorithmForTheSolutionOfTheKnapsackProblem}:
the quantum tree generator (QTG)-based search.
It can be understood as an enhanced version of quantum minimum finding~\cite{Durr1996AQuantumAlgorithmForFindingTheMinimum} in which Grover's algorithm~\cite{Grover1996AFastQuantumMechanicalAlgorithmForDatabaseSearch} is substituted by amplitude amplification~\cite{Brassard2002QuantumAmplitudeAmplificationAndEstimation} with the QTG as state preparation circuit.
Our main contribution is to embed the QTG into the QAOA framework, provide a theoretical convergence proof, and to benchmark the QTG-based QAOA against van Dam \textit{et al.}'s approach on the currently hardest 0-1-KP instances provided by Jooken \textit{et al.}~\cite{Jooken2022ANewClassOfHardProblemInstancesForTheKnapsackProblem} instance generator.

\section{Preliminaries}\label{section:Preliminaries}

\subsection{\label{subsection:TheKnapsackProblem}The knapsack problem}

Consider a knapsack with some capacity $c > 0$ as well as a list of $n$ items.
Each item $m$ is characterized by its profit value $v_{m} > 0$ and its weight $w_{m} > 0$.
The objective is to select a subset of items of maximum cumulative profit such that their cumulative weight does not exceed the knapsack's capacity.
By collecting the item profits and weights in lists $\bm{v} \coloneqq (v_{1}, \ldots, v_{n})$ and $\bm{w} \coloneqq (w_{1}, \ldots, w_{n})$, we address the instance, defined by these parameters, by $\kp$.
It has the following formulation as an Integer Linear Program (ILP):
\begin{align}\label{equation:KnapsackProblem}
\begin{split}
    \text{maximize } & \sum_{m = 1}^{n} v_{m} x_{m} \eqqcolon f(\bm{x}) \\
    \text{subject to } & \sum_{m = 1}^{n} w_{m} x_{m} \leq c \\
    \text{and } & x_{m} \in\{0, 1\}, \quad m = 1, \ldots, n.
\end{split}
\end{align}
The binary variable $x_{m}$ encodes the choice of either including item $m$ into the knapsack ($x_{m} = 1$) or omitting it ($x_{m} = 0$).

The 0-1-KP defined via \eqref{equation:KnapsackProblem} is shown to be NP-complete \cite{Karp1972ReducibilityAmongCombinatorialOptimizationProblems}, making it particularly interesting for potential quantum speed-ups.\footnote{Note however that there is a (classical) fully polynomial-time approximation scheme (FPTAS) for the 0-1-KP \cite{Ibarra1975FastApproximationAlgorithmsForTheKnapsackAndSumOfSubsetProblems}, approximating the optimal solution to a multiplicative factor $(1-\varepsilon)$ in time polynomial in $n$ and $1/\varepsilon$. Implementations are usually based on dynamic programming (DP) \cite{Bellman1957DynamicProgramming}, which is a general technique to solve a combinatorial optimization problem by reducing it to a recursive relation that can iteratively be computed. For 0-1-KP, this leads to the optimal solution in pseudo-polynomial time $\mathcal{O}(nc)$.}
Two common classical algorithms, both approximating the solution to \eqref{equation:KnapsackProblem} in linear time $\mathcal{O}(n)$, are the \emph{lazy greedy} (LG) and the \emph{very greedy} (VG) method.
Both variants assume the items to be sorted in descending order of their \emph{quality} $r_{m} \coloneqq v_{m} / w_{m}$.
Starting from the item of highest quality, the LG method includes items as long as the first does not fit into the knapsack anymore.
For later reference, we denote with $r_{\text{stop}}$ the quality of the first item which is not included by the LG method.
Similarly, the VG approach includes the items in the prescribed order, but does not stop at the first overshoot.
Non-fitting items are simply skipped, and the next ones are considered.
Therefore, VG always returns a (not necessarily proper) superset of the items included by LG.

\subsection{\label{subsection:QuantumTreeGenerator}Quantum tree generator}

Since the 0-1-KP is NP-complete \cite{Karp1972ReducibilityAmongCombinatorialOptimizationProblems}, methods based on Grover's algorithm \cite{Grover1996AFastQuantumMechanicalAlgorithmForDatabaseSearch}, and more generally amplitude amplification \cite{Brassard2002QuantumAmplitudeAmplificationAndEstimation}, promise an up to quadratic speed-up over a variety of classical approaches -- at least asymptotically.
In \cite{Wilkening2023AQuantumAlgorithmForTheSolutionOfTheKnapsackProblem}, this asymptotic dominance was recently calculated to already strike at instance sizes of a few hundreds of variables and thus within a practical domain.
The algorithmic procedure achieving the practical speed-up is a variant of \emph{Quantum Minimum Finding} (QMS) \cite{Durr1996AQuantumAlgorithmForFindingTheMinimum} with the uniform state preparation being replaced by the \emph{Quantum Tree Generator} (QTG) $\QTG$, operating on three registers: an $n$-qubit main (system) register encoding which items are included into the knapsack, a $\numbits{c}$-qubit capacity register which stores the remaining capacity of the knapsack, and a $\numbits{V}$-qubit profit register recording the cumulative profit of all included items.\footnote{$\numbits{a} = \lfloor \log_{2} a \rfloor + 1$ is the length of $a$'s binary representation.}
Here, $V$ is any upper bound on the maximum attainable profit.
The state preparation consists of $n$ subsequently applied layer unitaries $U_{m}$, one for each item $m$.
$U_{m}$ includes/excludes item $m$ in superposition if the remaining capacity covers item $m$'s weight and then updates the remaining capacity as well as the total profit on the branch, where the item has been included.
After applying all $n$ layer unitaries to the initial state $\ket{\bm{0}} \ket{c} \ket{0}$, the three registers hold a generally non-uniform superposition of all feasible item assignments together with their respective cumulative profit (see \autoref{fig:QTGExample} for a simple example for the QTG procedure).

\begin{figure*}[!ht]
    \centering
    \begin{minipage}{0.6\textwidth}
        \centering
        \tikzset{
          grow=down,
          level distance=2.8cm,
          sibling distance=3cm,
          edge from parent/.style={draw, -latex, thick},
          every node/.style = {font=\small, text centered},
          level 1/.style={sibling distance=9cm},
          level 2/.style={sibling distance=6cm},
          level 3/.style={sibling distance=3cm},
          node distance=1.5cm and 1.5cm
        }

        \begin{tikzpicture}[scale=0.65]
            \node[shape=rectangle, rounded corners,     draw, align=center,     top color=white, bottom color=blue!20] {$|000\rangle_S|3\rangle_{C}$}
            child {
              node[shape=rectangle, rounded corners,     draw, align=center,     top color=white, bottom color=blue!20] {$|000\rangle_S|3\rangle_{C}$} 
              child {
                node[shape=rectangle, rounded corners,     draw, align=center,     top color=white, bottom color=blue!20] {$|000\rangle_S|3\rangle_{C}$} 
                child { node[shape=rectangle, rounded corners,     draw, align=center,     top color=white, bottom color=blue!20] {$|000\rangle_S|3\rangle_{C}$} 
                  edge from parent
                  node[left, pos=0.5, above, sloped] {$\frac{1}{\sqrt{2}}$} }
                child { node[shape=rectangle, rounded corners,     draw, align=center,     top color=white, bottom color=blue!20] {$|001\rangle_S|2\rangle_{C}$} 
                  edge from parent
                  node[right, pos=0.5, above, sloped] {$\frac{1}{\sqrt{2}}$} }
                edge from parent
                node[left, pos=0.5, above, sloped] {$\frac{1}{\sqrt{2}}$}
              }
              child {
                node[shape=rectangle, rounded corners,     draw, align=center,     top color=white, bottom color=blue!20] {$|010\rangle_S|1\rangle_{C}$} 
                child { node[shape=rectangle, rounded corners,     draw, align=center,     top color=white, bottom color=blue!20] {$|010\rangle_S|1\rangle_{C}$} 
                  edge from parent
                  node[left, pos=0.5, above, sloped] {$\frac{1}{\sqrt{2}}$} }
                child { node[shape=rectangle, rounded corners,     draw, align=center,     top color=white, bottom color=blue!20] {$|011\rangle_S|0\rangle_{C}$} 
                  edge from parent
                  node[right, pos=0.5, above, sloped] {$\frac{1}{\sqrt{2}}$} }
                edge from parent
                node[right, pos=0.5, above, sloped] {$\frac{1}{\sqrt{2}}$}
              }
              edge from parent
              node[left, pos=0.5, above, sloped] {$\frac{1}{\sqrt{2}}$}
            }
            child {
              node[shape=rectangle, rounded corners,     draw, align=center,     top color=white, bottom color=blue!20] {$|100\rangle_S|0\rangle_{C}$}
              child { node[shape=rectangle, rounded corners,     draw, align=center,     top color=white, bottom color=blue!20] {$|100\rangle_S|0\rangle_{C}$}
                child { node[shape=rectangle, rounded corners,     draw, align=center,     top color=white, bottom color=blue!20] {$|100\rangle_S|0\rangle_{C}$} }
              }
              edge from parent
              node[right, pos=0.5, above, sloped] {$\frac{1}{\sqrt{2}}$}
            };
        \end{tikzpicture}
        \end{minipage}\hfill
        \begin{minipage}{0.4\textwidth}
        \begin{tcolorbox}[colframe=blue!20, colback=white, boxrule=0.5mm, enhanced, title=\textbf{\footnotesize Level 0}, attach boxed title to top center={yshift=-1mm}, coltitle=black, colbacktitle=blue!20, boxsep=1mm, left=0mm, right=0mm, top=0mm, bottom=0mm]
            \footnotesize The root represents the initial state having full capacity, since none of the three items are included.
        \end{tcolorbox}
        \vspace{-2mm}
        \begin{tcolorbox}[colframe=blue!20, colback=white, boxrule=0.5mm, rounded corners=south, enhanced, title=\textbf{\footnotesize Level 1}, attach boxed title to top center={yshift=-1mm}, coltitle=black, colbacktitle=blue!20, boxsep=1mm, left=0mm, right=0mm, top=0mm, bottom=0mm]
            \footnotesize This level splits into two branches: on the left nothing changed, while on the right the first item of weight $3$ is included.
        \end{tcolorbox}
        \vspace{-2mm}
        \begin{tcolorbox}[colframe=blue!20, colback=white, boxrule=0.5mm, rounded corners=south, enhanced, title=\textbf{\footnotesize Level 2}, attach boxed title to top center={yshift=-1mm}, coltitle=black, colbacktitle=blue!20, boxsep=1mm, left=0mm, right=0mm, top=0mm, bottom=0mm]
            \footnotesize On the left branch, each node further splits, while on the right nothing happens, since it already reached the maximum capacity.
        \end{tcolorbox}
        \vspace{-2mm}
        \begin{tcolorbox}[colframe=blue!20, colback=white, boxrule=0.5mm, rounded corners=south, enhanced, title=\textbf{\footnotesize Level 3}, attach boxed title to top center={yshift=-1mm}, coltitle=black, colbacktitle=blue!20, boxsep=1mm, left=0mm, right=0mm, top=0mm, bottom=0mm]
            \footnotesize The leaves show the feasible solutions with different amplitudes: the first four have an amplitude of $3 / \sqrt{2}$, while the last one of $1 / \sqrt{2}$.
        \end{tcolorbox}
    \end{minipage}
    \caption{Example of the quantum tree generator for the KP instance of three items with profits $\bm{v} = (4, 2, 1)$, weights $\bm{w} = (3, 2, 1)$ and maximum capacity $c = 3$. The quantum state $\ket{x_1, x_2, x_3}_S\ket{\tilde{c}}_{C}$ provides information about the inclusion or exclusion of the items via the system register $S$ ($x_i = 1$ if the item $i$ is included, else $0$) and indicates the remaining capacity $\Tilde{c}$ on capacity register $C$. The tree branches are equipped with the factors by which the child quantum states are rescaled.}
    \label{fig:QTGExample}
\end{figure*}

The inclusion/exclusion of item $m$ is executed via a controlled Ry-gate on the $m$-th qubit in the main register.
The introduced angle represents an adjustable bias towards either inclusion or exclusion;
they are, however, not treated as parameters to optimize over, but are updated according to fixed rules throughout the routine.
The control is over the capacity register and has the logical form ``$\tilde{c} \geq w_{m}$'', where $\tilde{c}$ is an integer represented by the current quantum state in the capacity register.
Only if this condition is fulfilled the Ry-gate is applied, otherwise the $m$-th qubit is left in the $\ket{0}$ state.
The updates of remaining capacity and profit are carried out via quantum Fourier transform (QFT) subtractors and adders, respectively.
$w_{m}$ is subtracted from $\tilde{c}$ in the capacity register and $v_{m}$ is added to $\tilde{V}$ in the profit register only if the state of the $m$-th qubit in the main register is $\ket{1}$.
In alignment with \cite{Wilkening2023AQuantumAlgorithmForTheSolutionOfTheKnapsackProblem}, we summarize these steps as $U_{m} = U_{m}^{3} U_{m}^{2} U_{m}^{1}$:
\begin{align}
\begin{split}
    U_{m}^{1} &= C_{\geq w_{m}}^{2}(\text{RY}_{m}(\theta)) \\
    U_{m}^{2} &= C_{m}^{1}(\text{SUB}_{w_{m}}) \\
    U_{m}^{3} &= C_{m}^{1}(\text{ADD}_{v_{m}})
\end{split}
\end{align}

\subsection{\label{subsection:QAOAGeneralDesign}QAOA: general design}

The few hundreds of logical qubits and long coherence times required for executing QTG-based search on non-trivial practical instances are out of reach for current quantum devices.
Hybrid algorithms such as the \emph{Quantum Approximate Optimization Algorithm} \cite{Farhi2014AQuantumApproximateOptimizationAlgorithm} / \emph{Quantum Alternating Operator Ansatz} \cite{Hadfield2019FromTheQuantumApproximateOptimizationAlgorithmToAQuantumAlternatingOperatorAnsatz} (QAOA) are candidates for supplying promising runtimes already on noisy quantum hardware.
While the original proposal is formulated for unconstrained optimization problems, subsequent refinements allow the QAOA to be applied to the general constrained case.

The three essential building blocks of the QAOA are a suitable initial state $\ket{\iota}$ and two parametrized quantum circuits: the \emph{phase separator} $\phasesep(\gamma)$ and the \emph{mixer} $\mixer(\beta)$.
Together with a prescribed \emph{depth} $q \in \N$, they define a parameterized class of states:
\begin{align}\label{equation:ParamertrizedStates}
    \ket{\bm{\beta}, \bm{\gamma}} \coloneqq \mixer(\beta_{q}) \phasesep(\gamma_{q}) \cdots \mixer(\beta_{1}) \phasesep(\gamma_{1}) \ket{\iota},
\end{align}
where $\bm{\beta} = (\beta_{1}, \ldots, \beta_{q})$ and $\bm{\gamma} = (\gamma_{1}, \ldots, \gamma_{q})$ combined constitute a set of $2 q$ real parameters.
Encoding the classical objective function $f$ associated with the optimization problem at hand into an objective Hamiltonian $H_{f}$ via
\begin{align}\label{equation:ObjectiveHamiltonian}
    H_{f} \coloneqq \sum_{\bm{x} \scriptin \{0, 1\}^{n}} f(\bm{x}) \ketbra{\bm{x}}{\bm{x}},
\end{align}
induces the standard phase separator
\begin{equation}\label{equation:PhaseSeparator}
    \phasesep^{f}(\gamma)=e^{-i \gamma H_{f}}.
\end{equation}
By construction, both $H_{f}$ and $U_{P}$ are diagonal in the computational basis.
This is desired for two reasons: First, it means that all Pauli strings $P_{k}$ in the Pauli decomposition $H_f=\sum_{k} \omega_k P_{k}$ mutually commute, which allows to implement \eqref{equation:PhaseSeparator} via
\begin{equation}
    \phasesep^f(\gamma) = e^{-i \gamma \sum_{k} \omega_{k} P_{k}} = \prod_{k} e^{-i \gamma \omega_{k} P_{k}}
\end{equation}
without any Trotter error.
The second reason will become clear later.
In contrast, the choice of the mixer $U_{M}$ is less straight-forward. As \eqref{equation:ParamertrizedStates} is derived from an approximated adiabatic evolution (see \cite{Farhi2000QuantumComputationByAdiabaticEvolution}), we have the freedom to pick any mixing Hamiltonian $H_{M}$ with highest-energy eigenstate $\ket{\iota}$ such that $U_{M}(\beta) = e^{-i \beta H_{M}}$.
Coming back to the objective Hamiltonian, \eqref{equation:ObjectiveHamiltonian} gives rise to the continuous function $F(\bm{\beta}, \bm{\gamma}) \coloneqq \expval{\bm{\beta}, \bm{\gamma}}{\, H_{f}\,}{\bm{\beta}, \bm{\gamma}}$.
Assuming that the ansatz is expressive enough to cover the entire quantum state space, the global maxima of $F$ exactly correspond to the (superpositions of) maxima of the classical objective function $f$ (compare \cite{Kossmann2022DeepCircuitQAOA}).
In any case, however, it is a most natural heuristic to maximize $F$ and expect the corresponding state $\ket{\bm{\beta}_{*}, \bm{\gamma}_{*}}$ to have sufficient overlap with computational basis states $\ket{\bm{x}}$ that correspond to high-valued bit strings.

For constrained optimization problems -- such as the 0-1-KP -- with feasible solution set $\feasset \subset \{0, 1\}^{n}$, the parameterized states that maximize $F$ typically lie outside of the \emph{feasible subspace} $\feasspace \coloneqq \spn \{\ket{\bm{x}} \defcolon \bm{x} \in S\}$ and would thus return infeasible bit strings upon measuring.
There are two main approaches addressing this issue: \emph{soft constraints} and \emph{hard constraints}.

Introducing soft constraints means suitably penalizing the objective value of infeasible solutions.
Namely, let $g : \{0, 1\}^{n} \rightarrow \{0, 1\}$ encode the classical constraints, that is $g(\bm{x}) = 1$ if and only if $\bm{x} \in \feasset$.
Introducing a sufficiently high \emph{penalty} $\alpha > 0$, replacing the objective function $f$ with $f - \alpha (1 - g)$, and constructing the corresponding objective Hamiltonian guarantees again that its maximum corresponds to feasible maxima of the classical objective function $f$.
However, choosing the penalty too high results in unfavorable optimization landscapes where the fine structure of the feasible subspace is entirely dominated by a hard cut-off to infeasible regions.
For maximization tasks such as 0-1-KP, there is a more natural, hyperparameter-free way to deal with solutions outside the feasible subspace: exchanging $f$ with $f g$, i.e., simply assigning the value zero to infeasible solutions.

\subsection{\label{subsection:QAOAStateOfTheArtForKnapsack}QAOA: state of the art for knapsack}

This approach of hyperparameter-free soft-constraints is also used in the most recent advancement regarding QAOA for 0-1-KP by van Dam \textit{et al.}~\cite{VanDam2021QuantumOptimizationHeuristicsWithAnApplicationToKnapsackProblems}.
While they alter the objective function to be $f g$, their phase separator remains the exponentiated original objective Hamiltonian \eqref{equation:PhaseSeparator}, that is $\phasesep^{f}$.\footnote{Note that it would also be possible to use some variation of $f$ in \eqref{equation:PhaseSeparator}, such as $fg$ or a threshold function, see \cite{Golden2021ThresholdBasedQuantumOptimization}.}
More specifically, replacing the binary variables $x_m$ by $(s_m + 1)/2$ with spin values $s_m=\pm 1$ in \eqref{equation:ObjectiveHamiltonian} yields the following simple form for the knapsack objective Hamiltonian:
\begin{align}\label{equation:KnapsackObjectiveHamiltonain}
    H_{f} = \sum_{m = 1}^{n} \frac{v_{m}}{2} \; \text{Z}_{m}.
\end{align}
Since all the terms in $H_{f}$ are single-qubit operations which mutually commute, the phase separator is non-entangling and can be implemented in depth one as
\begin{align}\label{equation:KnapsackPhaseSeparator}
    \phasesep(\gamma) = \bigotimes_{m = 1}^{n} \text{RZ}_{m}(2 v_{i}\gamma).
\end{align}
Constant terms in \eqref{equation:KnapsackObjectiveHamiltonain} could be omitted, as they only lead to global phases in \eqref{equation:KnapsackPhaseSeparator}.
Furthermore, since $H_{f}$ has an integer-valued spectrum, the exponential is $2\pi$-periodic in $\gamma$ which is why the parameter space is restricted to $\gamma \in [0, 2\pi)$.
For the initial state, van Dam \textit{et al.} propose the product state
\begin{align}\label{equation:CopulaInitialState}
    \ket{\iota} \hspace*{-0.75pt} = \hspace*{-0.75pt} \bigotimes_{m = 1}^{n} \ket{p_{m}} \; \text{with} \; \ket{p_{m}} \hspace*{-1pt} \coloneqq \hspace*{-1pt} \sqrt{1 - p_{m}} \ket{0} + \sqrt{p_{m}} \ket{1},
\end{align}
where
\begin{align}\label{equation:LogisticFunction}
    p_{m} = \frac{1}{1 + W e^{-k (r_{m} - r_{\text{stop}})}} \; \text{with} \; W = \sum_{m = 1}^{n} \frac{w_{m}}{c} - 1,
\end{align}
are classical probabilities determined by a hyperparameter $k \in \R^{+}$. Recall from \autoref{subsection:TheKnapsackProblem} that $r_{\text{stop}}$ is the quality (relative profit) of the first item that is not included anymore by LG.
The state in \eqref{equation:CopulaInitialState} can be prepared from the $\ket{\bm{0}}$ state via the depth-one circuit $\bigotimes_{m = 1}^{n} \text{RY}_{m}(2 \sin^{-1} (\sqrt{p_{m}}))$.
The logistic probability distribution produced by \eqref{equation:LogisticFunction} approaches the one given by LG for $k \to \infty$, and a uniform distribution for $k \to 0$.
Furthermore, van Dam \emph{et al.}introduce a family of copula mixers that aim at adjustable correlation patterns between different items.\footnote{They also introduce the non-entangling single-qubit hourglass mixer.
However, they also mention that since the knapsack phase separator is already non-entangling, the resulting QAOA variant can be efficiently simulated classically and shall therefore not count as a proper quantum algorithm.}
Since anti-correlating neighboring items consistently performed best in their numerical studies, we boil down their general construction to the anti-correlating case.
For two selected qubits $m$ and $m'$, define
the conditional probabilities
\begin{align}
    p_{m' | m} &= p_{m'} (1 - (1 - p_{m}) (1 - p_{m'})),  \label{equation:CorrelatedConditionalProbability1} \\
    p_{m' | \neg m} &= p_{m'} (1 + p_{m} (1 - p_{m'} )),
    \label{equation:CorrelatedConditionalProbability2}
\end{align}
based on the individual probabilities $p_{m}$ and $p_{m'}$ given by \eqref{equation:LogisticFunction}.
From these probabilities, construct the two-qubit rotational gate
\begin{align}\label{equation:TwoQubitCopulaRotation}
\begin{split}
    \text{R}(p_{m}, p_{m'}) &= \overline{\controlled}_{m} \Big(\text{RY}_{m'}\big(2 \sin^{-1}\hspace*{-3pt}\sqrt{p_{m' | \neg m}}\big)\Big) \\
    &\quad \times\, \controlled_{m} \Big(\text{RY}_{m'}\big(2 \sin^{-1}\hspace*{-3pt}\sqrt{p_{m' | m}}\big)\Big) \\
    &\quad \times\, \text{RY}_{m}\big(2 \sin^{-1}\hspace*{-3pt}\sqrt{p_{m}}\big).
\end{split}
\end{align}
The two-qubit copula mixer Hamiltonian is then given by
\begin{align}\label{equation:TwoQubitCopulaMixerHamiltonian}
    \text{Cop}_{m, m'} = \text{R}(p_{m}, p_{m'}) (\text{Z}_{m} + \text{Z}_{m'}) \text{R}(p_{m}, p_{m'})
\end{align}
which is exponentiated in order to yield the corresponding mixer:
\begin{align}\label{equation:TwoQubitCopulaMixer}
    U_{m, m'}^{\text{Cop}}(\beta) = \text{R}(p_{m}, p_{m'}) \text{RZ}_{m}(2 \beta) \text{RZ}_{m'}(2 \beta) \text{R}(p_{m}, p_{m'})^{\dagger}.
\end{align}
The (approximated) ring-copula mixer is a sequential mixer \cite{Hadfield2019FromTheQuantumApproximateOptimizationAlgorithmToAQuantumAlternatingOperatorAnsatz}, consisting of the product of all two-qubit copula mixers applied to neighboring items in a ring:
\begin{align}\label{equation:RingCopulaMixer}
    U_{\text{M}}^{\text{Cop}}(\beta) = \prod_{m = 1}^{n} U_{m, m + 1}^{\text{Cop}}(\beta).
\end{align}

\subsection{\label{subsection:QAOAGroverMixer}QAOA: Grover mixer}

In contrast to soft-constrained approaches like van Dam \textit{et al.}'s, enforcing hard constraints means actively restricting the parameterized class of states \eqref{equation:ParamertrizedStates} to the feasible subspace $\feasspace$.
This is typically achieved by choosing a feasible initial state $\ket{\iota} \in \feasspace$ as well as a feasibility-preserving mixer and phase separator.
This means that for all values for $\beta$ and $\gamma$, it holds that $\mixer(\beta)(\feasspace) \subseteq \feasspace$ and $\phasesep(\gamma)(\feasspace) \subseteq \feasspace$, respectively.
In this case, the objective function remains unchanged, as no infeasible states appear that would have to be penalized.

Now we can see the second advantage of choosing $H_f$ in \eqref{equation:ObjectiveHamiltonian} diagonal in the computational basis: As exponentiating the cost Hamiltonian in \eqref{equation:PhaseSeparator} preserves the diagonality, $\phasesep^f(\gamma)(\feasspace) \subseteq \feasspace$ immediately follows.

In contrast, the design of feasibility-preserving mixers is more involved.
There is no universal construction rule that inputs the classical constraints -- e.g. in the form of $g$ -- and outputs a suitable parameterized mixer $\mixer(\beta)$.
In practice, mixers are often constructed from symmetry structures of the underlying classical problems.
For example, the feasible set of the prominent traveling salesperson problem (TSP) can be described as the image of any fixed feasible solution under all elements of the symmetric group over $T$ elements, where $T$ is the number of cities to be visited (see \cite{Kossmann2023OpenShopSchedulingWithHardConstraints}).
The symmetric group, in turn, is generated by transpositions which can be straightforwardly quantized via swap gates. Exponentiation yields one possible TSP mixer choice (compare  \cite{Hadfield2019FromTheQuantumApproximateOptimizationAlgorithmToAQuantumAlternatingOperatorAnsatz}).

Employed symmetry aspects usually correspond to classical equality constraints.
For the 0-1-KP with its inequality constraint, a similar ansatz seems unlikely to exist.
However, the general design pattern of Grover mixers \cite{Baertschi2020GroverMixersForQAOAShiftingComplexityFromMixerDesignToStatePreparation} shifts the effort to a uniform, feasible state preparation circuit whose construction might not rely on symmetries:
Suppose that $\stateprep^{\text{Grov}}$ creates a uniform superposition of all feasible states, that is
\begin{align}\label{equation:UniformStatePreparation}
    \stateprep^{\text{Grov}} \ket{\bm{0}} = \ket{\feasset} \coloneqq \frac{1}{\sqrt{\abs{\feasset}}} \sum_{\bm{x} \scriptin \feasset} \ket{\bm{x}}.
\end{align}
The corresponding \emph{Grover mixer} is then defined via
\begin{align}\label{equation:GroverMixer}
\begin{split}
    \mixer^{\text{Grov}}(\beta) &\coloneqq e^{-i \beta \ketbra{\feasset}{\feasset}} \\
    &= \stateprep^{\text{Grov}} (\idty - (1 - e^{-i \beta}) \ketbra{\bm{0}}{\bm{0}}) (\stateprep^{\text{Grov}})^{\dagger},
\end{split}
\end{align}
where the last identity is obtained by writing out the exponential and leveraging the idempotence of the projector $\ketbra{\feasset}{\feasset}$. It also means that $\mixer^{\text{Grov}}$ is easy to implement once one has access to the state preparation routine $\stateprep^{\text{Grov}}$.
An instance of Grover mixer-QAOA (GM-QAOA) starts from the superposition state $\ket{\feasset}$ and employs the associated Grover mixer as well as the standard phase separator to build its parameterized class of states:
\begin{equation}\label{equation:GroverParametrizedStates}
    \ket{\bm{\beta}, \bm{\gamma}}^{\text{Grov}} = \left[\prod_{j=1}^q \mixer^{\text{Grov}}(\beta_j) \phasesep^f(\gamma_j) \right] \stateprep^{\text{Grov}} \ket{\bm{0}}.
\end{equation}

\section{Methods}\label{section:Methods}

\subsection{\label{subsection:QTGAGroverLikeMixer}QTG as Grover-like mixer}

We propose employing the quantum tree generator as basis for a Grover-like mixer. Note that the superposition created by the QTG is typically non-uniform, extending the concept of Grover mixers to more general state preparation routines.
In \autoref{subsection:ProofOfConvergence}, we argue that every superposition over feasible states with strictly positive amplitudes gives rise to a valid QAOA design.
Inspired by how Amplitude Amplification (AA) \cite{Brassard2002QuantumAmplitudeAmplificationAndEstimation} generalizes Grover's algorithm to nonuniform superpositions, we name our extended QAOA framework Amplitude Amplification-Mixer-QAOA (or AAM-QAOA). In analogy to \eqref{equation:GroverParametrizedStates}, the AAM-QAOA is characterized by states
\begin{equation}
    \ket{\bm{\beta}, \bm{\gamma}}^{\text{AA}} = \left[\prod_{j=1}^q \mixer^{\text{AA}}(\beta_j) \phasesep^f(\gamma_j) \right] \stateprep^{\text{AA}} \ket{\bm{0}}.
\end{equation}
For the quantum tree generator to be applicable in this context, we drop the profit register entirely and calculate the average profit during the intermediate measurements classically from the sampled bit strings instead.
Furthermore, in order to produce an equivalent of \eqref{equation:GroverMixer}, we slightly adjust the definition of the QTG to also start from the $\ket{\bm{0}}_S \ket{\bm{0}}_C$ state:
The capacity register is now also initialized in the $\ket{\bm{0}}_C$ state and item weights are added to the register's state rather than subtracted.
Accordingly, we change the control statement from ``$\tilde{c} \geq w_{m}$'' to ``$\tilde{c} \leq c - w_{m}$'', i.e. we check whether subtracting the $m$-th item's weight from the total capacity $c$ still covers the accumulated weight $\tilde{c}$.
Therefore, the left-hand side of this condition is again a quantum variable (the integer represented by the current quantum state of the capacity register) while the right-hand side is purely classical and can be precomputed. 
With this, the state preparation unitary $\stateprep^{\text{Grov}}$ can readily be replaced by $\QTG$. This creates a first representative of our AAM-QAOA framework with QTG mixer
\begin{equation}\label{equation:QTGMixer}
    \mixer^{\text{QTG}}(\beta) = \QTG \left[\idty - \left(1- e^{-i\beta}\right)\ketbra{\bm{0}}{\bm{0}}_S \otimes \ketbra{\bm{0}}{\bm{0}}_C\right] \QTG^{\dagger}.
\end{equation}

\subsection{\label{subsection:ParameterOptimization}Parameter optimization}

We follow the initialization strategy of van Dam \textit{et al.}~\cite{VanDam2021QuantumOptimizationHeuristicsWithAnApplicationToKnapsackProblems}, selecting initial parameters as the optimal candidates from a fine grid search to reduce the likelihood of convergence to local minima.
We then employ classical optimization techniques to fine-tune the initial parameters, specifically utilizing the constrained Powell algorithm as implemented in the nonlinear-optimization package (NLopt)~\cite{Johnson2007TheNLoptNonlinearOptimizationPackage}.
For $q = 1$ this reproduces van Dam \textit{et al.}'s approach.
For higher values of $q$, a joint initialization of all $2 q$ parameters via fine grid search with fixed granularity quickly becomes infeasible, as the number of points to be evaluated scales exponentially with $q$.
Instead, we follow an approach somewhat similar to the layer-wise optimization presented in~\cite{Lee2024IterativeLayerwiseTrainingForTheQuantumApproximateOptimizationAlgorithm}:
For all $i = 1, \ldots, q$, initialize the parameters $\beta_{i}$ and $\gamma_{i}$ in a two-dimensional fine grid search under fixed values for $\beta_{1}, \ldots, \beta_{i - 1}$ and $\gamma_{1}, \ldots, \gamma_{i - 1}$ and with all values $\beta_{i + 1}, \ldots, \beta_{q}$ and $\gamma_{i + 1}, \ldots, \gamma_{q}$ set to zero.
After determining the initial values for $\beta_{i}$ and $\gamma_{i}$, jointly optimize the parameters $\beta_{1}, \ldots, \beta_{i}$ and $\gamma_{1}, \ldots, \gamma_{i}$ while again keeping the remaining parameters fixed to zero.
This layer-wise initialization approach avoids the exponential cost of high-dimensional fine grid search;
yet, the subsequent optimization involves all parameters, reducing again the risk of converging to a local minimum.

In practice, the objective function $F(\bm{\beta}, \bm{\gamma})$ for the optimizer is implemented by sampling a fixed number of bit strings from the quantum computer after preparing the state $\ket{\bm{\beta}, \bm{\gamma}}$.
The obtained values are then plugged in into the classical objective function.
Taking the average of these values then yields an estimate for $F(\bm{\beta}, \bm{\gamma})$.
In our case, we additionally have to test for feasibility, since we wish to assign the value zero to infeasible states.
Therefore, each bit string is (again classically) tested for feasibility before calculating its objective value.

\subsection{\label{subsection:SimulationTechnique}Simulation technique}

There exist several suitable packages for simulating QAOA instances, e.g.\ \cite{Golden2023JuliQAOAFastFlexibleQAOASimulation}.
They are typically based on a state-vector-based simulation:
The current state of an $n$-qubit register is represented by a $2^{n}$-dimensional complex vector and the application of quantum gates is simulated by matrix-vector multiplication, usually in a suitable sparse format (such as in qHiPSTER~\cite{Smelyanskiy2016QHiPSTERTheQuantumHighPerformanceSoftwareTestingEnvironment}).
We follow this approach for the simulation of van Dam \textit{et al.}'s version of QAOA.
More precisely, the probabilities $p_{m}$ \eqref{equation:LogisticFunction} can be calculated classically, the approximated ring-copula mixer \eqref{equation:RingCopulaMixer} is implemented by subsequently executing the parameterized R-gates and RZ-gates in a qHiPSTER-like implementation.
In order to speed up the simulation, we employ another common trick:
We pre-compute the objective values of all $2^{n}$ bit strings and store them in an objective vector.
When simulating the application of $\phasesep^f(\gamma) = e^{-i \gamma H_{f}}$, we perform a element-wise product of the state vector with the element-wisely exponentiated objective vector.
This is highly parallelizable and therefore usually way faster than implementing the phase separator as a product of exponentiated Pauli strings.

For the simulation of the QTG-based AAM-QAOA, we do not follow a gate-wise strategy, but rather utilize aspects of the high-level simulator introduced in~\cite{Wilkening2023AQuantumAlgorithmForTheSolutionOfTheKnapsackProblem}.
More specifically, Wilkening \textit{et al.} developed a simulation technique specifically tailored to QTG-based search for the 0-1-KP.
Herein, they calculate the superposition created by the QTG via classically propagating the branching probabilities that arise from applying the $\text{RY}_{m}(\theta)$-gates.\footnote{
    More precisely, they calculate the probability distribution corresponding to the superposition created by the QTG.
    However, since all amplitudes in this superposition are guaranteed to be non-negative real numbers, we may simply take the respective (real) square root in order to obtain the correct amplitudes.
}
By adapting this subroutine of their simulator, we can therefore easily pre-compute the initial state $\ket{\text{KP}} \coloneqq \QTG \ket{\bm{0}}_S \ket{\bm{0}}_C$.
This further allows simulating the action of the QTG-based mixer thanks to the following identity:
\begin{align}\label{equation:QTGMixerSimulation}
\begin{split}
    \mixer^{\text{QTG}}(\beta) \ket{\psi} &= \big[\idty - (1 - e^{-i \beta}) \ketbra{\text{KP}}{\text{KP}}\big] \ket{\psi} \\
    &= \ket{\psi} - (1 - e^{-i \beta}) \braket{\text{KP} | \psi} \ket{\text{KP}},
\end{split}
\end{align}
which is an equivalent version of \eqref{equation:QTGMixer} obtained by writing out $\mixer^{\text{QTG}}=e^{-i\beta \ketbra{\text{KP}}{\text{KP}}}$.
Evaluating $\braket{\text{KP} | \psi}$ as well as the subsequent subtraction in \eqref{equation:QTGMixerSimulation} are both again highly parallelizable and therefore yield a comparatively fast simulation.
Since we are, by design, guaranteed to stay within the feasible subspace, the simulation volume can be restricted.
Accordingly, we cut all entries from the objective vector which correspond to infeasible solutions, and remove all zero entries within $\ket{\text{KP}}$, which are precisely the amplitudes of the infeasible states.
Restricting the simulation to the feasible subspace of course destroys the tensor product structure which is used in e.g.\ qHiPSTER to speed up the simulation of elementary gates.
Fortunately, as we have just detailed, we can simulate all components of the QTG-QAOA with much faster methods that do not rely on exploiting any tensor product structure.

\section{Results}\label{section:Results}

\subsection{\label{subsection:ProofOfConvergence}Proof of convergence}

Quantitative performance guarantees of variational methods like the QAOA are notoriously difficult to derive.
However, qualitative reachability questions can often be answered with substantially less effort.
Concretely: Given an initial state $\ket{\iota}$, a mixer $B$, and a phase separator $C$, do they define a family of parameterized quantum circuits that, given a sufficiently large depth $p$, allows for sampling an optimal solution with almost certainty?
This notion of QAOA convergence is captured in \cite[Theorem 12]{Binkowski2024ElementaryProofOfQAOAConvergence}.
In the following, we validate that GM-QAOA and its generalization to positive non-uniform superpositions of all feasible states -- AAM-QAOA -- indeed fulfill the sufficient criteria for convergence.
These criteria are phrased in terms of the Hamiltonians which are exponentiated in order to yield the phase separator and mixer, respectively.

First, the phase separator Hamiltonian $C$ merely has to be diagonal in the computational basis such that the eigenspace of the restriction $C\vert_{\feasspace}$ corresponding to its largest eigenvalue matches
\begin{align}\label{equation:OptimalSubspace}
    \langle \feasset_{\max}\rangle \coloneqq \spn\{\ket{\bm{x}} \defcolon \bm{x} \text{ is an optimal solution}\},
\end{align}
which is trivially fulfilled for the choice of $C = H_{f}$.

Second, the mixer Hamiltonian $B$ has to be feasibility-preserving, i.e. $B(\feasspace) \subseteq \feasspace$, and $B\vert_{\feasspace}$, when considered as a matrix in the computational basis, has to be irreducible with non-negative entries only.
A matrix is called irreducible if it does not leave any non-trivial coordinate subspace, i.e.\ a linear subspace spanned by a proper subset of the basis chosen to represent the matrix, invariant.

Let $\ket{\phi}$ be an arbitrary state with expansion in the computational basis
\begin{align}\label{equation:NonUniformSuperposition}
    \ket{\phi} \coloneqq \sum_{\bm{x} \in \{0, 1\}^{n}} \phi_{\bm{x}} \ket{\bm{x}},
\end{align}
and consider the projector $\ketbra{\phi}{\phi}$ as candidate for a mixer Hamiltonian.
Furthermore, let $\ket{\psi} \in \feasspace$ be an arbitrary state from the feasible subspace;
then the application of $\ketbra{\phi}{\phi}$ yields
\begin{align}\label{equation:ProjectionApplication}
\begin{split}
    \ketbraket{\phi}{\phi}{\psi} &= \sum_{\bm{x} \in \{0, 1\}^{n}} \sum_{\bm{y} \in \{0, 1\}^{n}} \sum_{\bm{z} \in \feasset} \phi_{\bm{x}}^{\vphantom{*}} \overline{\phi}_{\bm{y}} \psi_{\bm{z}}^{\vphantom{*}} \ketbraket{\bm{x}}{\bm{y}}{\bm{z}} \\
    &= \sum_{\bm{x} \in \{0, 1\}^{n}} \phi_{\bm{x}} ^{\vphantom{*}}\Bigg(\sum_{\bm{z} \in \feasset} \overline{\phi}_{\bm{z}} \psi_{\bm{z}}^{\vphantom{*}}\Bigg) \ket{\bm{x}}.
\end{split}
\end{align}
The result is again in $\feasspace$ if and only if either the summation of $\overline{\phi}_{\bm{z}} \psi_{\bm{z}}^{\vphantom{*}}$ over $\bm{z} \in \feasset$ yields zero or all the coefficients $\phi_{\bm{x}}$ for $\bm{x} \notin \feasset$ are zero.
The first case can only hold for a generic $\ket{\psi} \in \feasspace$ if all the coefficients $\phi_{\bm{z}}^{\vphantom{*}}$ for $\bm{z} \in \feasset$ are zero.
Therefore, we conclude that
\begin{align}
    \ketbra{\phi}{\phi}\!(\feasspace) \subseteq \feasspace \iff \ket{\phi} \in \feasspace \text{ or } \ket{\phi} \in \feasspace^{\perp}. 
\end{align}
However, the second option is not of interest for us since it implies that $\ketbra{\phi}{\phi}\vert_{\feasspace} = 0$ which fails to be irreducible (as it leaves every subspace invariant).
Therefore, we are left with superpositions of feasible computational basis states.

Furthermore, the matrix elements of the restriction $\ketbra{\phi}{\phi}\vert_{\feasspace}$ are given by
\begin{align}
    \braket{\bm{x} \ketbra{\phi}{\phi} \bm{y}} = \phi_{\bm{x}}^{\vphantom{*}} \overline{\phi}_{\bm{y}},\quad \bm{x}, \bm{y} \in \feasset. 
\end{align}
These are collectively non-negative if and only if all the amplitudes $\phi_{\bm{x}}$, $\bm{x} \in \feasset$, are non-negative.

Lastly, we consider the irreducibility of the matrix representation of $\ketbra{\phi}{\phi}\vert_{\feasspace}$.
First, assume that all amplitudes $\phi_{\bm{x}}$ are non-zero. 
Let $\{0\} \subsetneq \langle R \rangle \subsetneq \feasspace$ be an arbitrary non-trivial coordinate subspace of $\feasspace$ with $\emptyset \neq R \subsetneq \feasset$ being a proper, non-empty subset of all feasible solutions.
By assumption, there exist $\bm{y} \in R$ and $\bm{z} \in \feasset \setminus R$ such that both $\phi_{\bm{y}} \neq 0$ and $\phi_{\bm{z}} \neq 0$.
From \eqref{equation:ProjectionApplication} we infer that $\ketbraket{\phi}{\phi}{\bm{z}}$ has a component of $\phi_{\bm{y}} \phi_{\bm{z}} \neq 0$ in direction of $\ket{\bm{z}} \notin \langle R\rangle$, i.e.\ the coordinate subspace $\langle R \rangle$ is not invariant under $\ketbra{\phi}{\phi}$.
Since $R$ was chosen arbitrarily, this establishes irreducibility.
Conversely, if there exists a $\bm{z} \in \feasset$ with $\phi_{\bm{z}} = 0$, the non-trivial coordinate subspace $\spn\{\ket{\bm{z}}\}$ would be left invariant by $\ketbra{\phi}{\phi}$ as all of its elements are collinear with $\bm{z}$ which is being mapped to zero according to \eqref{equation:ProjectionApplication}.
In summary, the projection $\ketbra{\phi}{\phi}$ fulfills all the mixer criteria if and only if all the amplitudes $\phi_{\bm{x}}$ in \eqref{equation:NonUniformSuperposition} vanish for $\bm{x} \notin \feasset$ and are strictly positive for $\bm{x} \in \feasset$.

Third, the initial state $\ket{\iota}$ has to be the largest energy state of the restricted mixer Hamiltonian $B\vert_{\feasspace}$.
Choosing $B = \ketbra{\phi}{\phi}$, AAM-QAOA prescribes to take $\ket{\iota} = \ket{\phi}$.
As a projection, $\ketbra{\phi}{\phi}\vert_{\feasspace}$ only has the eigenvalues zero and one, with $\ket{\phi}$ being an eigenstate to the (highest) eigenvalue one.

We summarize that AAM-QAOA fulfills all requirements listed in \cite[Theorem 12]{Binkowski2024ElementaryProofOfQAOAConvergence} which thereby implies that, given a sufficiently large depth $p$ and suitable parameter values $\bm{\beta}$ and $\bm{\gamma}$, the state $\ket{\bm{\beta}, \bm{\gamma}}$ will be arbitrarily close to the optimal solution space $\langle \feasset_{\max}\rangle$ and with this yield an optimal solution with almost certainty.

\subsection{\label{subsection:NumericalExperiments}Numerical experiments}

We benchmark the QTG-based QAOA as representative of the AAM-QAOA family against the state-of-the-art Copula approach by van Dam \textit{et al.}~\cite{VanDam2021QuantumOptimizationHeuristicsWithAnApplicationToKnapsackProblems} on the hardest 0-1-KP instances that can currently be found, created by the instance generator of Jooken \textit{et al.}~\cite{Jooken2022ANewClassOfHardProblemInstancesForTheKnapsackProblem}.
In particular, they are more complex to solve than the classes of instances constructed by Pisinger~\cite{Pisinger2005WhereAreTheHardKnapsackProblems}, on which the numerical studies in~\cite{VanDam2021QuantumOptimizationHeuristicsWithAnApplicationToKnapsackProblems} are based.
The high-level technique described in \autoref{subsection:SimulationTechnique} allows simulating of our algorithm on instances with up to $n = 34$ items.
The standard gate-based simulation, in contrast, limits the tractable problem size to $n = 20$ when running the Copula-QAOA.
On the other hand, this upper bound results from a trade-off between the number of items and the depth $q$, i.e., number of angles to be optimized.
While the number of elapsed cycles is independent of the problem size in van Dam \textit{et al.}'s ansatz (and the gate count only shows a moderate scaling), this is not the case for our approach.
The resources required by the QTG-QAOA are mainly driven by the demands of the quantum tree generator.
In \cite{Wilkening2023AQuantumAlgorithmForTheSolutionOfTheKnapsackProblem}, Wilkening \textit{et al.} conducted an extensive analysis, showing how their method scales in terms of qubits, gates and cycles.
Provided gate execution times on the specific hardware at hand, the latter can be used as a proxy for the algorithm's runtime.
This is based on the assumption of all-to-all qubit connectivity and suitable parallelization capabilities, meaning that disjoint gates, acting on different qubits, may be executed in parallel, constituting a single cycle on a QPU.
Irrespective of the chosen configuration of phase separator, mixer and initial state, \eqref{equation:ParamertrizedStates} implies an upper bound
\begin{equation}\label{equation:CycleCountQAOA}
    \cyclec_{\text{QAOA}} = q (\cyclec_{\text{P}} + \cyclec_{\text{M}}) + \cyclec_{\text{SP}}
\end{equation}
on the QAOA cycle demand, where $\cyclec_{\text{P}}$ and $\cyclec_{\text{M}}$ denote the cycle counts of phase separator and mixer, respectively, and $\cyclec_{\text{SP}}$ accounts for the initial preparation of $\ket{\iota}$.
For our algorithm with $\ket{\iota} = \ket{\text{KP}}$, this means one additional application of $\QTG$.
On the other hand, the initial state \eqref{equation:CopulaInitialState} proposed by van Dam \textit{et al.} gives $\cyclec^{\text{Cop}}_{\text{SP}} = 1$ for the Copula-QAOA (see \autoref{subsection:QAOAStateOfTheArtForKnapsack}).
Using the standard phase separator $\phasesep^{f}$ from \eqref{equation:PhaseSeparator}, $\cyclec^{f}_{\text{P}} = 1$ as argued in \eqref{equation:KnapsackPhaseSeparator} for 0-1-KP.
The cycle cost of the QTG mixer \eqref{equation:QTGMixer} is given by
\begin{equation}
    cyclec^{\text{QTG}}_{\text{M}} = 2 \cyclec_{\QTG} + \cyclec_{\text{MC-P}}
\end{equation}
where the number of cycles $\cyclec_{\text{MC-P}}$, required to implement the multi-controlled phase gate sandwiched by the QTG in \eqref{equation:QTGMixer}, depends on the chosen decomposition method; here a standard Toffoli cascade (see, e.g., \cite[Fig. 4.10]{NielsenChuang2010QuantumComputationAndQUantumInformation})
\begin{align}
\begin{split}
    \cyclec_{\text{MC-P}} &= \cyclec_{\text{TOF}} + 1 = 2 \numbits{(\qubitc_{\text{control}} - 1)} + 1 \\
    &= 2 \numbits{(n + \numbits{c} - 2)} + 1
\end{split}
\end{align}
where the number of control qubits $\qubitc_{\text{control}}$ is given by the combined length of the system and the capacity register minus one.
In line with \cite{VanDam2021QuantumOptimizationHeuristicsWithAnApplicationToKnapsackProblems}, the Copula mixer \eqref{equation:RingCopulaMixer} can be implemented within a constant amount of cycles, only alternating between even and odd problem sizes:
\begin{equation}
    \cyclec^{\text{Cop}}_{\text{M}} = \cyclec^{\text{Cop}}_2 \cdot \begin{cases}
        2 & \; \text{if} \; n \; \text{is even} \\ 3 & \; \text{if} \; n \; \text{is odd}
    \end{cases}
\end{equation}
where
\begin{equation}
    \cyclec^{\text{Cop}}_{2} = 2 \cyclec_{\text{R}} + 1 = 7,
\end{equation}
with $\cyclec^{\text{Cop}}_{2}$ and $\cyclec_{\text{R}}$ referring to \eqref{equation:TwoQubitCopulaMixer} and \eqref{equation:TwoQubitCopulaRotation}, respectively.

\begin{figure}[!t]
    \centering
    \includegraphics[width=0.48\textwidth]{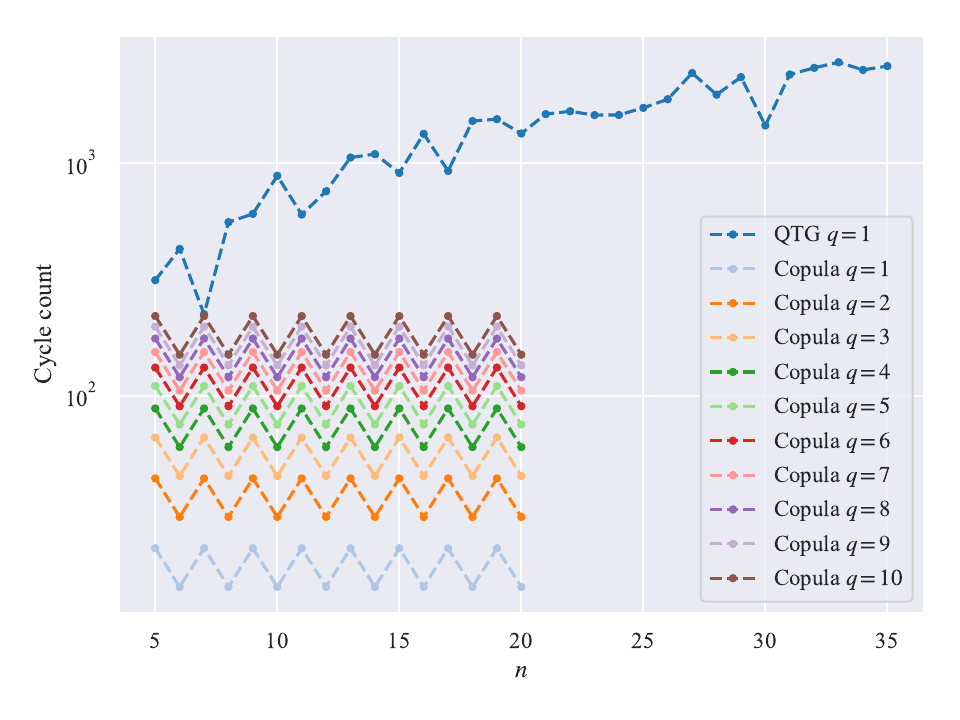}
    \caption{Cycle counts for the QTG-QAOA and Copula-QAOA implementations. Due to the more expensive QTG state preparation, larger depths $q$ are allowed for the Copula-based approach. The number of elapsed QPU cycles is calculated under the assumption of all-to-all qubit connectivity and that gates acting on disjoint qubits may be executed in parallel. For a fixed value $q$, the cycle count of the Copula-QAOA does not depend on the concrete problem size $n$, but alternates between even and odd values for $n$. In contrast, the QTG-approach's cycle demands overall increase with the number of items. Irregular dips result from lower item weights and profits which, in turn, allow for shorter addition circuits within the QTG. For a fixed value of $n$, the required resources grow linearly with $q$, independent of the QAOA type, but only observable for the Copula-based method.}
    \label{fig:cycle-counts}
\end{figure}

\begin{figure}[!t]
    \centering
    \includegraphics[width=.48\textwidth]{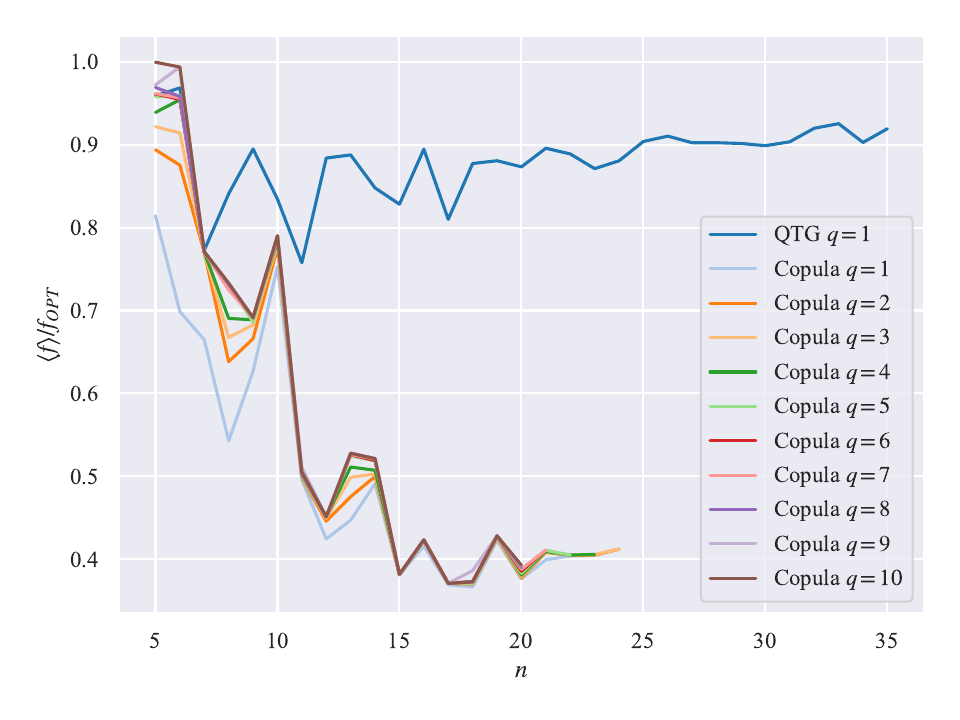}
    \caption{Approximation ratios of the QTG-QAOA and Copula-QAOA implementations. Due to the more expensive QTG state preparation, larger depths $q$ are allowed for the Copula implementation. For increasing values up to the maximum investigated $q=10$, convergence among the Copula curves can be observed, especially for larger problem sizes $n$. Only for tiny instances could Copula outperform the QTG-based implementation, which stabilizes at about 0.9 approximation ratio for large problem sizes. The former settles at values around 0.4 after a rapid decline with small intermediate peaks.}
    \label{fig:approximation-ratio}
\end{figure}

\autoref{fig:cycle-counts} shows how \eqref{equation:CycleCountQAOA} evaluates on the generated instances for both approaches.
Due to the more expensive QTG state preparation, we can afford more layers of phase separator and mixer applications in the Copula-QAOA.
A viable assessment of our algorithm's performance would ideally necessitate equal resource demands.
Comparing the respective cycle costs on the generated instances, we find that a $q$ value between 101 and 102 would at worst be needed in the Copula-based approach to catch up with our algorithm's implementation at lowest possible $q=1$ (for $n = 18$).
However, since \autoref{fig:approximation-ratio} shows good convergence for depth values iterated between $q = 1$ and $q = 10$, especially for larger problem instances, such prohibitive orders of magnitude can be safely dismissed.

We numerically demonstrate the superiority of our approach over the state-of-the-art Copula-QAOA by investigating the global approximation ratio achieved by both algorithms (see \autoref{fig:approximation-ratio});
this is, the final expectation value returned by the QAOA in proportion to the objective function value of the optimal solution found by COMBO.
All subroutines -- of both QTG and Copula -- are devoid of any kind of randomization, avoiding the need for expensive averaging corrections.
We observe that, while results of the QTG-QAOA are confined in a vertical interval between 0.75 and 0.97 approximation ratio, the Copula-QAOA experiences a rapid decline from QTG-like values at $n = 5$ to ratios ranging from 0.54 to 0.73 for eight items.
At these problem sizes, a clear Copula-specific improvement can be identified for growing depths, with $q = 1$ performing particularly worse than the others.
The Copula results merge to an intermediate peak for all $q$ values at $n = 10$, confirming the behavior found in~\cite{VanDam2021QuantumOptimizationHeuristicsWithAnApplicationToKnapsackProblems}.
After that, they collectively drop to values below 0.53 -- and settle around 0.4 approximation ratio.
The QTG curve, on the other hand, rises again after fluctuating heavily until instance sizes of about 18 items.
More specifically, the approximation ratios achieved by our implementation seem to stabilize around a value of approximately 0.9.
Only for tiny 0-1-KP instances was the Copula ansatz able to beat our algorithm.

\begin{figure}[!t]
    \centering
    \includegraphics[width=.48\textwidth]{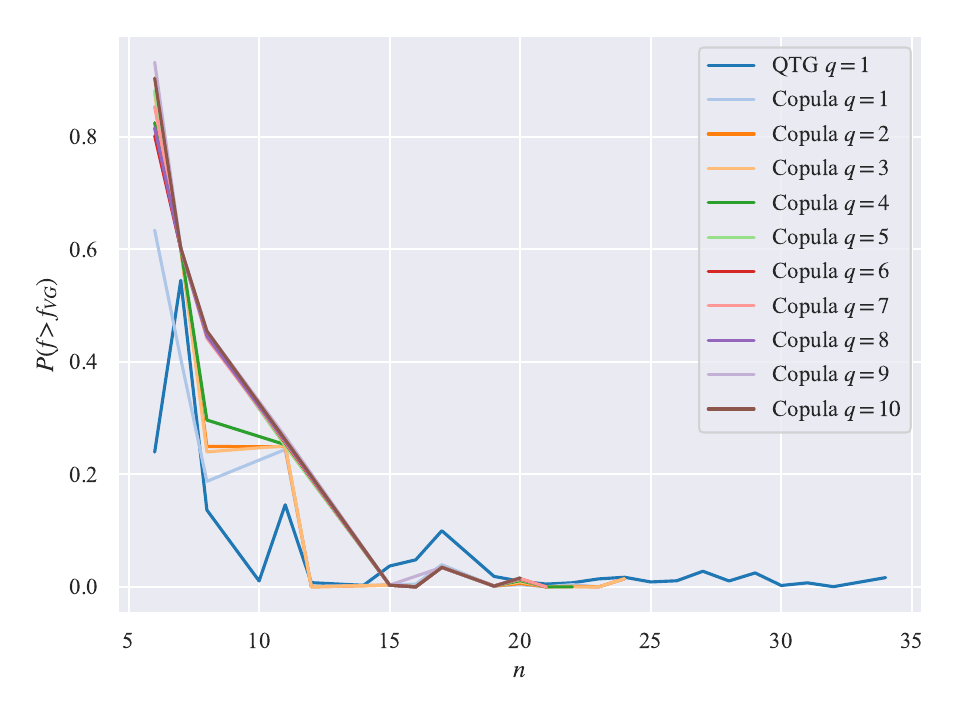}
    \caption{Probability of finding a feasible solution $x$ with objective value $f(x)$ greater than the very greedy (VG) solution value $f_{\text{VG}}$. Both implementations feature a reasonable probability for small instances, which drops significantly for increasing problem size. For instances with $n \geq 15$, large overlap among all Copula curves can be observed. Increasing the depth beyond $q = 5$ appears to have no effect at all. The Copula-QAOA performs better initially, but shows a quicker decline, even for larger numbers of angles. Instances where the VG heuristic found the optimal solution have been excluded.}
    \label{fig:prob}
\end{figure}

\autoref{fig:prob} illustrates the probability of both approaches to outperform the VG heuristic (see \autoref{subsection:TheKnapsackProblem}), which is obtained by summing up the individual probabilities of all (feasible) solution states inside the final state-vector, whose profits are larger than the VG solution.
While we observe reasonable outcomes for small instances consisting of up to 11 items, the probability nearly vanishes completely for larger problem sizes.
The last bit of a chance of beating VG fades out after $n = 16$, where QTG rears up considerably to a 0.1-sized peak for the final time.
Here, the slight resurrection of the Copula-QAOA is rather negligible, even for larger numbers of mixer/phase-separator layers and angles, sealing the faster Copula shortfall compared to our approach.
It should be noted, however, that it achieves significantly better probability values between five and ten items compared to the QTG-based AAM-QAOA, with the global maxima being around 0.9 and 0.55, respectively.
These Copula results again confirm the findings reported in~\cite{VanDam2021QuantumOptimizationHeuristicsWithAnApplicationToKnapsackProblems}.
Remarkably, increasing the depth beyond $q = 5$ in van Dam \textit{et al.}'s ansatz appears to have no measurable effect on the probability of beating VG.
Instances where the VG heuristic found the optimal solution have been excluded from \autoref{fig:prob} to circumvent ambiguity as, in this case, the probability of ending up better than VG vanishes independently of the respective QAOA outcomes.

\section{Conclusion}\label{section:Conclusion}

In this work, we have introduced a QAOA method for the 0-1 knapsack problem that incorporates hard-constraints through the use of the Quantum Tree Generator (QTG)~\cite{Wilkening2023AQuantumAlgorithmForTheSolutionOfTheKnapsackProblem} as basis for a Grover-like mixer~\cite{Baertschi2020GroverMixersForQAOAShiftingComplexityFromMixerDesignToStatePreparation}.
The QTG-based QAOA may be seen as a representative of the more general family of Amplitude Amplification-Mixer-QAOA (AAM-QAOA) where the necessity of a uniform state preparation is dropped.
We benchmarked our algorithm against the state-of-the-art Copula approach developed by van Dam \textit{et al.}~\cite{VanDam2021QuantumOptimizationHeuristicsWithAnApplicationToKnapsackProblems}, utilizing the most challenging 0-1 knapsack problem instances generated by Jooken \textit{et al.}~\cite{Jooken2022ANewClassOfHardProblemInstancesForTheKnapsackProblem}.

Our numerical results demonstrate that our method consistently outperforms the Copula-based ansatz in terms of the approximation ratio as independent measure.
Specifically, only in very small instances does the Copula-QAOA show better performance compared to our QTG-based implementation.
We also evaluate the probability of both methods outperforming the classical (very) greedy approach and observed that, for larger problem sizes, this probability nearly vanishes.
Although the Copula-QAOA exhibits superior performance for smaller instances compared to the QTG-QAOA, it rapidly declines in efficacy as problem size increases, even when massively extending the circuit depth and, thereby, the number of variational parameters.

At least for arbitrarily large circuit depths and the right parameters, we can prove that AAM-QAOA, and thus the QTG-based QAOA as a special case, will find the global optimal solution.
Our proof utilizes the conditions for QAOA convergence as being phrased in~\cite{Binkowski2024ElementaryProofOfQAOAConvergence}.
This purely theoretical result may be of independent interest.
At the same time, it gives very little implications for the practical performance of QAOA at finite depth.

Despite achieving a better QAOA method with improved approximation ratios compared to Copula-QAOA, our findings suggest that the classical greedy approach remains more effective overall for large and complex problem instances.
Further research may explore enhancements to quantum algorithms that can rival classical heuristics in a broader range of problem sizes.

\section*{Acknowledgment}

This work was supported by the Quantum Valley Lower Saxony and the BMBF projects ProvideQ and QuBRA.
We thank Jan Rasmus Holst, Tobias J.\ Osborne and Timo Ziegler for helpful discussions.

\noindent\textbf{Data and code availability statement.}
The depicted data and the used code can be found at \url{https://github.com/Damuna/QAOA-with-Grover-mixer}.

\IEEEtriggeratref{23}
\bibliographystyle{IEEEtran}
\bibliography{IEEEabrv.bib,main.bib}

\end{document}